# Multi-qubit quantum phase gates based on surface plasmons of a nanosphere

**Jun Ren[1], Jun Yuan[1,2] and Xiangdong Zhang[1*]**

[1]School of Physics, Beijing Institute of Technology, 100081, Beijing, China
[2]Department of Physics, Beijing Normal University, Beijing 100875, China
\* *Corresponding author, E-Mail address:zhangxd@bit.edu.cn*

## Abstract

The Dicke subradiance and superradiance resulting from the interaction between surface plasmons of a nanosphere and an ensemble of quantum emitters have been investigated by using a Green's function approach. Based on such an investigation, we propose a scheme for a deterministic multi-qubit quantum phase gate. As an example, two-qubit, three-qubit and four-qubit quantum phase gates have been designed and analyzed in detail. Phenomena due to the presence of losses in the metal are discussed. The potential application of present phenomena to the quantum-information processing is anticipated.

**OCIS codes:** (270.5585); (240.6680)

# 1. INTRODUCTION

In the last decades, many efforts have been undertaken to study quantum gates, because they play an important role in quantum-information processing and quantum-computing networking [1]. There are a large number of proposals for two-qubit gate operations in ion trap, cavity-QED system, NMR, quantum dots and superconducting charge qubits [3-7]. Multiqubit-controlled quantum gates are also very useful in the construction of quantum-computing networks, implementing quantum-error-correction protocols and quantum algorithms. Therefore, some investigations have also been performed in several physical schemes such as three-qubit Toffoli gates with neutral atoms in an optical lattice [8], hybrid atom-photon qubit via a cavity-QED system [9], multiqubit unitary gate using adiabatic passage with a single-mode optical cavity [10], n-qubit-controlled phase gate with superconducting quantum-interference devices coupled to a resonator [11] and so on.

On the other hand, the surface plasmons have been subject of recent studies because they have opened up new possibilities in many fields [12] such as realizing nanoscale photonic components [13], overcoming the diffraction limit of light [14, 15], producing subdiffraction nanolaser [16, 17] or enhanced fluorescence [18–20]. It has been proposed as an efficient single-photon generator [21] as well as a single-photon transistor [22], strong emitter-plasmon coupling has been experimentally shown using quantum dots [23], and single plasmons along the wire have been detected by using N-V centers as emitters [24]. The spontaneous formation of qubit entanglement mediated by plasmons has also been proposed [25] and wave-particle duality of surface plasmons has been verified in experiments [26].

Recently, quantum emitters coupled to surface plasmons of the nanowire and cooperative emission of light by dipoles near a metal nanosphere have been investigated, subradiance and superradiance resulting from the plasmon-mediated interaction have been demonstrated [27, 28]. Based on surface plasmons of the nanowire, a scheme for a deterministic two-qubit quantum gate has been proposed [27]. In this work, we explore the possibility to realize multi-qubit quantum gates based on the surface plasmons of a nanosphere and an ensemble of quantum emitters.

# 2. THE GENERATION OF SUPERRADIANCE AND SUBRADIANCE STATES

We consider N two-level atoms located in the vicinity of a metal sphere as shown in Fig. 1. Let us further assume that the atoms are sufficiently far from each other, so that the interatomic Coulomb interaction can be ignored. Under the electric-dipole and rotating wave approximations, the Hamiltonian of the system can be expressed as [27, 29]

$$\hat{H} = \int d^3\vec{r} \int_0^\infty d\omega \hbar\omega \hat{\vec{f}}^\dagger(\vec{r},\omega)\hat{\vec{f}}(\vec{r},\omega) + \sum_A \frac{1}{2}\hbar\omega_A \hat{\sigma}_{Az} - \sum_A [\hat{\sigma}_A^\dagger \hat{\vec{E}}^{(+)}(\vec{r}_A)\vec{d}_A + H.c.], \qquad (1)$$

where $\vec{r}_A$ and $\omega_A$ represent the position operator and transition frequency of the atom, respectively. Here $\vec{d}_A$ represents the dipole moment, $\hat{\sigma}_A^\dagger = |e\rangle_{A\ A}\langle g|$, $\hat{\sigma}_A = |g\rangle_{A\ A}\langle e|$ and $\hat{\sigma}_{Az} = |e\rangle_{A\ A}\langle e| - |g\rangle_{A\ A}\langle g|$ are Pauli operators for the two-level atoms, $|g\rangle_A$ and $|e\rangle_A$ represent

the ground and excited states of the atom, respectively, $\hat{\vec{f}}^{\dagger}(\vec{r},\omega)$ and $\hat{\vec{f}}(\vec{r},\omega)$ are bosonic field operators which play the role of the fundamental variables of the electromagnetic field and the medium. The electric-field operator is expressed in terms of $\hat{\vec{f}}(\vec{r},\omega)$ as [29]

$$\hat{\vec{E}}^{(+)}(\vec{r}) = i\sqrt{\frac{\hbar}{\pi\varepsilon_0}} \int_0^{\infty} d\omega \frac{\omega^2}{c^2} \int d^3\vec{r}\,' \sqrt{\varepsilon_I(\vec{r}\,',\omega)} \vec{\vec{G}}(\vec{r},\vec{r}\,',\omega) \hat{\vec{f}}(\vec{r}\,',\omega), \qquad (2)$$

where $\vec{\vec{G}}(\vec{r},\vec{r}\,',\omega)$ is the classical Green tensor and $\varepsilon_I(\vec{r}\,',\omega)$ is the complex permittivity. For a single-quantum excitation, the system wavefunction at time t can be written as

$$\begin{aligned}|\psi(t)\rangle &= \sum_A C_{U_A}(t) e^{-i(\omega_A - \bar{\omega})t} |U_A\rangle|\{0\}\rangle \\ &+ \int d^3\vec{r} \int_0^{\infty} d\omega \left[ C_{Li}(\vec{r},\omega,t) e^{-i(\omega - \bar{\omega})t} |L\rangle|\{1_i(\vec{r},\omega)\}\rangle \right], \end{aligned} \qquad (3)$$

where $\bar{\omega} = \frac{1}{2}\sum_A \omega_A$, $|U_A\rangle$ represents the excited upper state of the Ath atom and all the other atoms are in the lower state, $|\{0\}\rangle$ is the vacuum state of the medium, $|L\rangle$ is the atomic lower state and here $|\{1_i(\vec{r},\omega)\}\rangle = \hat{f}_i^{\dagger}(\vec{r},\omega)|\{0\}\rangle$ is the state of the medium which is excited in a single-quantum Fock state, $C_{U_A}(t)$ and $C_{Li}(\vec{r},\omega,t)$ are the probability amplitudes of the states $|U_A\rangle|\{0\}\rangle$ and $|L\rangle|\{1_i(\vec{r},\omega)\}\rangle$. If we assume that the excitation initially resides in one of the atoms, that means $C_{U_A}(0) = 1$ and $C_{Li}(\vec{r},\omega,0) = 0$, from the Schrodinger equation, we obtain the following integro-differential equations [29]:

$$\dot{C}_{U_A}(t) = \sum_{A'} \int_0^t dt' K_{AA'}(t,t') C_{U_{A'}}(t'), \qquad (4)$$

$$K_{AA'}(t,t') = -\frac{1}{\hbar\pi\varepsilon_0} \int_0^{\infty} d\omega [\frac{\omega^2}{c^2} e^{-i(\omega-\omega_A)t} e^{i(\omega-\omega_{A'})t'} \vec{d}_A \,\text{Im}\,\vec{\vec{G}}(\vec{r}_A,\vec{r}_{A'},\omega) \vec{d}_{A'}]. \qquad (5)$$

We assume all the atoms have the same transition frequencies and they follow the same decay laws, consider the symmetry of Green tensor, from Eq.(5) we obtain $K_{AB}(t-t') = K_{BA}(t-t')$ for any two atoms, and $K_{AA}(t,t') = K_{BB}(t,t') = K_{CC}(t,t') = \cdots \equiv K(t,t')$ for all the atoms. In order to solve the Eq. (4), we take a new basis in the following form:

$$|i\rangle = \sum_{A'}^{N} x_{A'} |U_{A'}\rangle. \qquad (6)$$

The corresponding probability amplitude is expressed as:

$$C_i = \sum_A^N x_{A'} C_{U_{A'}} \tag{7}$$

and

$$\dot{C}_i(t) = \sum_{A''}^N \int_0^t dt' [\sum_A^N x_{A'} K_{A'A''}(t-t')] C_{U_{A''}}(t') . \tag{8}$$

If we let $\dfrac{\sum_A^N x_{A'} K_{A'A''}}{x_{A''}} = k$, that is, $\dfrac{\sum_A x_{A'} K_{A'A}}{x_A} = \dfrac{\sum_A x_{A'} K_{A'B}}{x_B} = \dfrac{\sum_A x_{A'} K_{A'C}}{x_C} = \cdots = k$, then

$$\begin{vmatrix} K-k & K_{BA} & K_{CA} & \cdots & K_{NA} \\ K_{AB} & K-k & K_{CB} & \cdots & K_{NB} \\ K_{AC} & K_{BC} & K-k & \cdots & K_{NC} \\ \vdots & \vdots & \vdots & \ddots & \vdots \\ K_{AN} & K_{BN} & K_{CN} & \cdots & K-k \end{vmatrix} = 0 . \tag{9}$$

Every eigenvalue for k corresponds to the quantum state $|i_n\rangle \begin{pmatrix} x_A & x_B & x_C & \cdots & x_N \end{pmatrix}$. Once $k$ has been obtained, the motion equation of quantum states for the probability amplitudes can be expressed as

$$\dot{C}_i(t) = \int_0^t k C_i(t') dt' . \tag{10}$$

Then

$$C_i(t) = e^{-(\Gamma/2 + i\delta)t} \tag{11}$$

with

$$\delta = \frac{1}{x_{A''}} \sum_A^N x_{A'} \delta_{A'A''} \quad \text{and} \quad \Gamma = \frac{1}{x_{A''}} \sum_A^N x_{A'} \Gamma_{A'A''}, \tag{12}$$

where

$$\Gamma_{AA'} = \frac{2\omega_A^2}{\hbar \varepsilon_0 c^2} \vec{d}_A \operatorname{Im} \ddot{G}(\vec{r}_A, \vec{r}_{A'}, \omega_A) \vec{d}_{A'} , \tag{13}$$

$$\delta_{AA'} = \frac{P}{\pi \hbar \varepsilon_0} \int_0^\infty d\omega \frac{\omega^2}{c^2} \frac{\vec{d}_A \operatorname{Im} \ddot{G}(\vec{r}_A, \vec{r}_{A'}, \omega_A) \vec{d}_{A'}}{\omega - \omega_A} . \tag{14}$$

If we consider that the direction of the dipole moment $\vec{d}_A$ is in parallel of that for $\vec{r}$, from the above equations and the classical Green tensor according to Ref. [29], we can obtain

$$\begin{aligned} \frac{\Gamma_{AB}}{\Gamma_0} = 6\pi \{ & \operatorname{Re} \sum_{n=1}^\infty \sum_{m=-n}^n n(n+1) \frac{h_n(k_1 r) j_n(k_1 r)}{(k_1 r)^2} Y_{nm}(\theta_A, \varphi_A) Y^*_{nm}(\theta_{A'}, \varphi_{A'}) + \\ & \operatorname{Re} \sum_{n=1}^\infty \sum_{m=-n}^n n(n+1) (\frac{h_n(k_1 r)}{(k_1 r)})^2 Y_{nm}(\theta_A, \varphi_A) Y^*_{nm}(\theta_{A'}, \varphi_{A'}) b_{nm} \} \end{aligned} \tag{15}$$

Here $r$ represents the distance between the atom and the center of the sphere, $k_1 = n_1 k_0 = \sqrt{\varepsilon_1} k_0 = \sqrt{\varepsilon_1} \frac{\omega}{c}$, $b_{nm}$ is the scattering coefficient. The $\theta_A$ and $\varphi_A$, $\theta_{A'}$ and $\varphi_{A'}$ are the angle positions for atom $A$ and $A'$, respectively. And $\Gamma_0 = \frac{\omega^3 d_A^2}{3\hbar\pi\varepsilon_0 c^3}$ represents the decay rate of an atom in free space. $\Gamma_{AA'}$ is the contribution which describes the effect of atoms upon each other. Thus, $\Gamma_{AA'}/\Gamma_0$ is a function as relative positions and transition frequencies of atoms. Before investigating on the physical properties of $\Gamma_{AA'}/\Gamma_0$, we discuss the spontaneous emission of an excited atom placed near a metal sphere. Figure 2 (b) shows the single-atom decay rate as a function of the frequency. Here the radius of the sphere is taken as a=20nm and the atoms are placed at r=25nm from the center of the sphere. For the metallic sphere, we use the frequency-dependent dielectric constant, $\varepsilon(\omega) = 1 - [\omega_p^2 / \omega(\omega + i\gamma)]$. Here following Ref. [30], we have chosen $\omega_p = 6.18 eV$ and $\gamma = 0.05 eV$. Some enhanced peaks of the spontaneous decay are observed at $\omega = 0.55\omega_p$, $0.625\omega_p$ and $0.652\omega_p$. These frequencies correspond to the eigenmodes of plasmon excitation for n=1, 2, 3 as shown in Fig.2 (a). This means that the plasmon excitations of the metal sphere produce important effect on the spontaneous emission of the atom.

If we place many emitters around the metal sphere, all at the same distance from the surface, close enough to couple to the surface plasmon mode, they are expected to have a long range interaction mediated by the plasmons. For example, if we arrange the positions of atom A $(r_A, \theta_A, \varphi_A)$ and B $(r_B, \theta_B, \varphi_B)$ as shown in Fig.3(a), this is equal to assuming the atom A being located at $\theta_A = \pi/2, \varphi_A = 0.0$, and the position of atom B is $\theta_B = \pi/2, \varphi_B = \varphi$. Thus, the normalized cross relaxation rate $\Gamma_{AB}$ can be expressed as [29]

$$\Gamma_{AB}/\Gamma = \{6\pi[\text{Re}\sum_{n=1}^{\infty}\sum_{m=-n}^{n} n(n+1)\frac{h_n(k_1 r)j_n(k_1 r)}{(k_1 r)^2} Y_{nm}(\frac{\pi}{2},0.0)Y^*_{nm}(\frac{\pi}{2},\varphi) + \text{Re}\sum_{n=1}^{\infty}\sum_{m=-n}^{n} n(n+1)(\frac{h_n(k_1 r)}{(k_1 r)})^2 Y_{nm}(\frac{\pi}{2},0.0)Y^*_{nm}(\frac{\pi}{2},\varphi)b_{nm}]\Gamma_0\}/\Gamma \quad (16)$$

where $\Gamma$ is the decay of single atom, from Ref.[31] it is expressed as

$$\Gamma = (1 + \frac{3}{2}\sum_{n=1}^{\infty} n(n+1)(2n+1)\text{Re}\{b_n \frac{[h_n(k_1 r)]^2}{(k_1 r)^2}\})\Gamma_0 \quad . \quad (17)$$

Figures 3 shows the calculated results of $\Gamma_{AB}/\Gamma$ as a function of the angle $\varphi$. The parameters of the metal sphere are taken identical with those in Fig.2. Here the atoms are placed at r=25nm as shown in Fig.3(a). The solid line, dashed line and dotted line correspond to the cases with $\omega = 0.55\omega_p, 0.625\omega_p$ and $0.652\omega_p$, respectively. We find that $\Gamma_{AB}/\Gamma$ for various frequencies exhibit different oscillatory behavior with the change of angle. At some angles for the certain frequencies, $\Gamma_{AB}/\Gamma$ posses some maximum values, corresponding to predominantly superradiant states while the smaller decay rates are those of predominantly subradiant states [27]. Such a phenomenon is similar to the case of the metal nanowire as has been discussed in Ref.[27]. The advantage to use the metal sphere instead of the metal nanowire is that it is convenient to realize superradiant and subradiant states of many atoms by using spherical symmetry. Furthermore, it is convenient to construct multi-qubit quantum phase gates.

### 3. MULTI-QUBIT QUANTUM PHASE GATE

As a possible application, we can construct a phase gate by exploiting the large difference in the decay rates between the above superradiant and subradiant states [27]. We assume that only the $|g\rangle \to |e\rangle$ transition of each emitter is coupled to the single plasmonic excited mode of the metal sphere. Because the emitters are sufficiently far from each other, they can individually be addressed by lasers coupling the $|g\rangle \to |e\rangle$ transition. Assume that we can ignore the decay of the excited level $|e\rangle$ in the $|g\rangle - |e\rangle$ two level system like the described in Ref.[27]. In this case it is well known that a $2\pi$ pulse, forcing the system to do a full Rabi oscillation, gives an additional $\pi$ phase to the atomic system [27]. On the other hand if the decay rate of the excited state is much stronger than the resonant Rabi frequency, the driving field cannot induce Rabi oscillations between the two levels, and the drive merely introduces a scattering rate. Therefore, the subradiant and superradiant effects are of decisive importance to realize quantum phase gates. The plasmonic excited modes around the metal sphere provide the possibility to construct such subradiant and superradiant states. In the following, we discuss how to realize two-qubit, three-qubit and four-qubit quantum phase gates by using subradiant and superradiant states of the metal sphere.

#### 3.1. Two-qubit quantum phase gate

If we only consider the level scheme of the two-atom system as has been described in Ref.[27], Eq. （9）is reduced as

$$\begin{vmatrix} K-k & K_{AB} \\ K_{BA} & K-k \end{vmatrix} = 0 \quad , \tag{18}$$

where $K_{AB}(t-t') = K_{BA}(t-t')$ and $\Gamma_{AB} = \Gamma_{BA}$, then the eigenvalues are $k = K \pm K_{AB}$ and the corresponding quantum states can be expressed as：

$$\begin{aligned} |1\rangle &= 1/\sqrt{2}(|U_A\rangle - |U_B\rangle) \\ |2\rangle &= 1/\sqrt{2}(|U_A\rangle + |U_B\rangle) \end{aligned} \quad . \tag{19}$$

Then the probability amplitudes of quantum states are written as

$$C_1(t) = e^{(-\Gamma_1/2 + i\delta_1)t} C_1(0)$$
$$C_2(t) = e^{(-\Gamma_2/2 - i\delta_1)t} C_1(0) \qquad (20)$$

with

$$\Gamma \equiv \Gamma_{AA}, \delta \equiv \delta_{AA}$$
$$\Gamma_1 = \Gamma - \Gamma_{AB}, \delta_1 = \delta + \delta_{AB} \qquad (21)$$
$$\Gamma_2 = \Gamma + \Gamma_{AB}, \delta_2 = \delta - \delta_{AB}$$

As the transition probability of the atom is taken as $\omega_A = 0.625\omega_p$, which corresponds to the surface polariton mode n=2 in Fig.2(b), the atom A is put on the position with $\theta_A = \frac{\pi}{2}$ and $\varphi_A = 0$, and the atom B on the position with $\theta_B = \frac{\pi}{2}$ and $\varphi_B = \pi$ as shown in Fig.5 (d), we obtain $\Gamma_{AB} = \Gamma$, $\Gamma_2 = 2\Gamma > \Gamma$ and $\Gamma_1 = 0$. This means that the spontaneous radiation for $|1\rangle$ is zero (subradiant state), and it is maximum for $|2\rangle$ (superradiant state). The ground and excited state of atoms A and B are coupled by resonant classical drive fields with a strength parametrized by the Rabi frequencies $\Omega_A$ and $\Omega_B$, i.e., $H_i = \Omega_i(|g\rangle_i\langle e| + |e\rangle_i\langle g|)$ $(i = A, B)$. These fields couple the states $|gg\rangle, |eg\rangle, |ge\rangle$ and $|ee\rangle$. Thus, the effective coupling strengths for $|1\rangle$ and $|2\rangle$ are written as

$$\Omega_1 = \frac{1}{\sqrt{2}}(\Omega_A + \Omega_B), \qquad (22)$$

$$\Omega_2 = \frac{1}{\sqrt{2}}(\Omega_A - \Omega_B). \qquad (23)$$

Assume $\Omega_A = \Omega_B$, $\Omega_1 = \sqrt{2}\Omega_A$ and $\Omega_2 = 0$. Choosing the driving strength $\Omega_1$ to be in between the two decay rates $\Gamma_2 \gg \Omega_1 \gg \Gamma_1$ ensures that transition to the double excited state $|ee\rangle$ is blocked by the strong decay $\Gamma_2$, whereas we can perform a $2\pi$ Rabi oscillation on the two-level system given by $|gg\rangle$ and $|1\rangle$ as shown in Fig.4 (a). This thus ensures that we achieve a phase change in $\pi$ on the $|gg\rangle$ state. Due to the strong decay $\Gamma_2$, the transition $|gg\rangle \to |2\rangle$ is also blocked. As for the transitions $|sg\rangle \to |se\rangle$ and $|gs\rangle \to |es\rangle$ as shown in Fig.4 (b), there is essentially only a single atom interaction with the sphere since there is no coupling to the state $|s\rangle$. The excited state therefore always has a fast decay $\Gamma \sim \Gamma_1 / 2$ such that the driving is too weak to excite the atoms. Furthermore, the state $|ss\rangle$ is completely unaffected by the classical light pulses. Thus, for the initial state

$$|\psi_{\text{intinal}}\rangle = \frac{1}{2}(|ss\rangle + |sg\rangle + |gs\rangle + |gg\rangle), \qquad (24)$$

through the following transformation under the certain drive field,

$$|ss\rangle \to |ss\rangle; |sg\rangle \to |sg\rangle; |gs\rangle \to |gs\rangle; |gg\rangle \to -|gg\rangle, \qquad (25)$$

it becomes

$$|\psi_{\text{ideal}}\rangle = \frac{1}{2}(|ss\rangle + |sg\rangle + |gs\rangle - |gg\rangle). \qquad (26)$$

This means that the two-qubit quantum phase gate has been implemented.

The efficiency of the phase gate can be described by fidelity. The fidelity of the gate is defined as the square of the absolute value of the overlap between the desired final atomic state and the one we get in the end, that is [27]

$$F = |\langle \psi_{\text{ideal}} | \psi \rangle|^2 = \langle \psi_{\text{ideal}} | \psi \rangle \langle \psi | \psi_{\text{ideal}} \rangle. \qquad (27)$$

In fact, such a result can be also estimated by using scaling analysis from the following expression [27].

$$F = 1 - \sqrt{\frac{\Gamma_1}{\Gamma}}. \qquad (28)$$

For example, the fidelity at the above case with r=25nm is about 71%. The value of fidelity changes with the distance between two atoms and the metal sphere. The red line in Fig. 5 (b) describes the fidelity as a function of the distance between atoms and the sphere center. The maximum of the fidelity is about 83%, which appears at r=43.5nm. This means that the fidelity can be optimized by taking suitable distance between two atoms and the metal sphere, which is similar to the case of two-quibt phase gate based on the metal wire as has been described in Ref.[27].

In fact, the value of fidelity strongly depend on the absorption loss of the metal material, it can be improved by coating gain medium on the metal sphere to overcome the absorption loss. Figure 5 (a) shows the calculated results of $\Gamma_{AB} / \Gamma$ as a function of the angle when the metal sphere is coated with different dielectric gain materials as shown in Fig.5(c). Here the atoms are placed at r=30nm. The blue line represents the calculated result for the coated metal sphere with the dielectric constant of the gain material $\varepsilon' = 1.2 - 0.01i$. Here the radius of the coated sphere is taken as b=24nm, and the metal core is identical with the metal sphere in Fig.3. The green line is the corresponding result for the coated metal sphere with $\varepsilon' = 1.2 - 0.05i$. Comparing them with the case without coating (red line), we find that the value of $\Gamma_{AB} / \Gamma$ increases a little. However, the fidelity increases largely with introducing coating layer. The blue line and green line in Fig.5 (b) display the fidelity as a function of the distance between atoms and the sphere center for two kinds of coating, respectively. The fidelity can reach 95% at r=30nm for the coated metal sphere with $\varepsilon' = 1.2 - 0.05i$.

Such a method is more effective for the metal sphere with strong absorption. Figure 6 shows the fidelity as a function of the distance between atoms and the center of aluminum sphere being coated with different dielectric gain materials. For the aluminum, we have chosen $\omega_p = 2.27 \times 10^4$ THz and $\gamma = 0.05\omega_p$ [32]. The radii of aluminum core and coated sphere are still taken as a=20nm and b=24nm, respectively. The solid line corresponds to the case without coating, the dashed line and dotted line correspond to the aluminum sphere coated by $\varepsilon' = 1.2 - 0.1i$ and $\varepsilon' = 1.2 - 0.16i$, respectively. Here $\omega = 0.57\omega_p$ is taken, which corresponds to one of the surface polariton modes. It is seen clearly that the fidelities corresponding to various atom positions are below 50% due to the strong absorption loss. For example, it is only 38.7% for the case with r=30nm. However, it becomes 55.2% and 74.5% when the aluminum sphere is coated by $\varepsilon' = 1.2 - 0.1i$ and $\varepsilon' = 1.2 - 0.16i$, respectively.

### 3.2. Three-qubit quantum phase gate

Similar to the two-quibit case, we can also realize three-qubit quantum phase gate based on the surface plasmon of the nanosphere. We consider three two-level atoms located in the vicinity of the metal sphere. Assuming $K_{AB} = K_{BC} = K_{CA}$ and $\Gamma_{AB} = \Gamma_{BC} = \Gamma_{AC}$ in solving Eq.(9), we obtain quantum states under the eigenvalues $k = K - K_{AB}$ and $k = K + 2K_{AB}$ in the following form:

$$|3\rangle = \frac{1}{\sqrt{3}}(|U_A\rangle + |U_B\rangle + |U_C\rangle)$$
$$|4\rangle = \frac{1}{\sqrt{6}}(2|U_A\rangle - |U_B\rangle - |U_C\rangle) \quad (29)$$
$$|5\rangle = \frac{1}{\sqrt{6}}(-|U_A\rangle + 2|U_B\rangle - |U_C\rangle)$$

with the probability amplitudes of quantum states being written as

$$C_i(t) = e^{(-\Gamma_i/2 + i\delta_i)t} C_i(0); i = 3,4,5, \quad (30)$$

where

$$\Gamma_3 = \Gamma + 2\Gamma_{AB}, \delta_3 = \delta + 2\delta_{AB}$$
$$\Gamma_{4(5)} = \Gamma - \Gamma_{AB}, \delta_{4(5)} = \delta - \delta_{AB} \quad (31)$$

If we let $\Gamma = -2\Gamma_{AB}$, then $\Gamma_3 = 0$ and $\Gamma_{4(5)} = \frac{3}{2}\Gamma > \Gamma$. For such a case, the spontaneous radiation for $|3\rangle$ is zero (subradiant state), and it is maximum for $|4\rangle$ and $|5\rangle$ (superradiant states). Such a case can be realized by putting three atoms on the positions with $\theta_A = \theta_B = \theta_C = 0.3\pi$, $\varphi_A = 0$, $\varphi_B = \frac{2\pi}{3}$ and $\varphi_C = \frac{4\pi}{3}$ as shown in the inset of Fig.9(a). Here the transition probabilities of the atoms are also taken as $\omega_A = 0.625\omega_p$, which also corresponds to the surface polariton mode n=2 in Fig.2(b). Thus, the effective coupling strengths of the resonant classical drive fields for $|3\rangle$, $|4\rangle$ and $|5\rangle$ are written as

$$\Omega_3 = \frac{1}{\sqrt{3}}(\Omega_A + \Omega_B + \Omega_C), \tag{32}$$

$$\Omega_4 = \frac{1}{\sqrt{6}}(2\Omega_A - \Omega_B - \Omega_C), \tag{33}$$

$$\Omega_5 = \frac{1}{\sqrt{6}}(-\Omega_A + 2\Omega_B - \Omega_C). \tag{34}$$

Let $\Omega_A = \Omega_B$, then $\Omega_3 = \sqrt{3}\Omega_A$ and $\Omega_{4(5)} = 0$. Choosing the driving strength $\Omega_3$ to be in between the two decay rates $\Gamma_{4(5)} \gg \Omega_3 \gg \Gamma_3$ ensures that the transitions $|ggg\rangle \to |4\rangle \ or \ |5\rangle$ are blocked by the strong decay $\Gamma_4$ and $\Gamma_5$, whereas the transition $|ggg\rangle \to |3\rangle$ appears as shown in Fig.7 (a). At the same time, a phase change in $\pi$ on the $|ggg\rangle$ state has been obtained. As for the transitions $|gss\rangle, |sgs\rangle, |ssg\rangle \to |ess\rangle, |ses\rangle, |sse\rangle$ as shown in Fig.6 (b), the corresponding decay is $\Gamma$ and the transitions are blocked. For the initial state $|sgg\rangle$, there is only interaction of two atoms with the sphere as shown in Fig.7 (c). In such a case, $|s1\rangle = \frac{1}{\sqrt{2}}(|sge\rangle + |seg\rangle)$ and $|s2\rangle = \frac{1}{\sqrt{2}}(|sge\rangle - |seg\rangle)$, the corresponding decay rates for them are $\Gamma_1 = \frac{1}{2}\Gamma$ and $\Gamma_3 = \frac{3}{2}\Gamma$, respectively. So, the transitions $|sgg\rangle \to |s1\rangle \ or \ |s2\rangle$ are also blocked due to fast decay. For the initial states $|gsg\rangle$ and $|ggs\rangle$, the situation is similar to that of $|sgg\rangle$. In addition, the transitions related to the state $|sss\rangle$ do not occur due to absence of the interaction between the state $|s\rangle$ and surface modes of the metal sphere. Thus, the three-qubit quantum phase gate to perform the transition $|ggg\rangle \to -|ggg\rangle$ can be only implemented in the above system by the certain drive field.

### 3.3. Four-qubit quantum phase gate

The above method can also apply to the design of four-qubit quantum phase gate. Assume $K_{AC} = K_{BD} = K_{AB} = K_{BC} = K_{CD} = K_{DA}, \Gamma_{AB} = \Gamma_{BC} = \Gamma_{CD} = \Gamma_{DA} = \Gamma_{AD} = \Gamma_{AC}$ in solving Eq.(9) with four atoms, we obtain quantum states under the eigenvalues $k = K - K_{AB}$ and $k = K + 3K_{AB}$ in the following form:

$$|6\rangle = \frac{1}{\sqrt{4}}(|U_A\rangle + |U_B\rangle + |U_C\rangle + |U_D\rangle)$$

$$|7\rangle = \frac{1}{\sqrt{12}}(3|U_A\rangle - |U_B\rangle - |U_C\rangle - |U_D\rangle)$$

$$|8\rangle = \frac{1}{\sqrt{12}}(-|U_A\rangle + 3|U_B\rangle - |U_C\rangle - |U_D\rangle)$$

$$|9\rangle = \frac{1}{\sqrt{12}}(-|U_A\rangle - |U_B\rangle + 3|U_C\rangle - |U_D\rangle)$$

(35)

with probability amplitudes of quantum states being written as

$$C_i(t) = e^{(-\Gamma_i/2 + i\delta_i)t} C_i(0); i = 6,7,8,9 ,$$

(36)

where

$$\Gamma_6 = \Gamma + 3\Gamma_{AB}, \delta_6 = \delta + 3\delta_{AB}$$
$$\Gamma_{7(8,9)} = \Gamma - \Gamma_{AB}, \delta_{7(8,9)} = \delta - \delta_{AB}.$$

(37)

If four atoms are put on the acme of the positive quadrilateral, any two of the atoms have the angle relations: $\theta_A = \theta_B = 0.392\pi$ and $\varphi = \frac{2}{3}\pi$ as shown in the inset of Fig.9 (b), we obtain $\Gamma_6 = 0$ and $\Gamma_{7(8,9)} = \frac{4}{3}\Gamma > \Gamma$ under assuming $\Gamma_{AB} = -\Gamma/3$. In such a case, the spontaneous radiation for $|6\rangle$ is zero (subradiant state), and it is maximum for $|7\rangle, |8\rangle, |9\rangle$ (superradiant states). Here the transition probabilities of atoms are identical with those in Fig.4. The effective coupling strengths of resonant classical drive fields for $|6\rangle, |7\rangle, |8\rangle$ and $|9\rangle$ are expressed as

$$\Omega_6 = \frac{1}{\sqrt{4}}(\Omega_A + \Omega_B + \Omega_C + \Omega_D),$$

(38)

$$\Omega_7 = \frac{1}{\sqrt{12}}(3\Omega_A - \Omega_B - \Omega_C - \Omega_D),$$

(39)

$$\Omega_8 = \frac{1}{\sqrt{12}}(-\Omega_A + 3\Omega_B - \Omega_C - \Omega_D),$$

(40)

$$\Omega_9 = \frac{1}{\sqrt{12}}(-\Omega_A - \Omega_B + 3\Omega_C - \Omega_D). \tag{41}$$

As $\Omega_A = \Omega_B$, $\Omega_6 = 2\Omega_A$ and $\Omega_{7(8,9)} = 0$. Similar to the above cases, choosing the driving strength $\Omega_6$ to be in between the two decay rates $\Gamma_{7(8,9)} \gg \Omega_6 \gg \Gamma_6$ ensures that the transition $|gggg\rangle \to |6\rangle$ appears, whereas the transitions $|gggg\rangle \to |7\rangle, |8\rangle, |9\rangle$ are blocked by the strong decay $\Gamma_7, \Gamma_8, \Gamma_9$ as shown in Fig.8 (a), a phase change in $\pi$ on the $|gggg\rangle$ state has been given. However, the transitions related to the state $|ssss\rangle$ do not occur. The transitions $|gsss\rangle, |sgss\rangle, |ssgs\rangle, |sssg\rangle \to |esss\rangle, |sess\rangle, |sses\rangle, |ssse\rangle$ as shown in Fig.8 (b) are also blocked by the decay $\Gamma$. For the initial state $|ssgg\rangle$, only interaction of two atoms with the drive field as shown in Fig.8 (c), we let $|ss1\rangle = \frac{1}{\sqrt{2}}(|ssge\rangle + |sseg\rangle)$ and $|ss2\rangle = \frac{1}{\sqrt{2}}(|ssge\rangle - |sseg\rangle)$, the corresponding decay rates for them are $\Gamma_1 = \frac{2}{3}\Gamma$ and $\Gamma_2 = \frac{4}{3}\Gamma$, respectively. .So, the transitions described in Fig.8 (c) can not occur, which is similar to the case in Fig.7 (c). As for the initial state $|sggg\rangle$, there are interactions of three atoms with the drive field as shown in Fig.8 (d). If we assume

$$|s3\rangle = \frac{1}{\sqrt{3}}(|sgge\rangle + |sgeg\rangle + |sgge\rangle) \quad , \quad |s4\rangle = \frac{1}{\sqrt{6}}(2|sgge\rangle - |sgeg\rangle - |sgge\rangle) \quad \text{and}$$

$$|s5\rangle = \frac{1}{\sqrt{6}}(-|sgge\rangle + 2|sgeg\rangle - |sgge\rangle), \text{ the corresponding decay rates, } \Gamma_3 = \frac{1}{3}\Gamma \text{ and}$$

$\Gamma_{4,5} = \frac{4}{3}\Gamma$, are still big. The transitions $|sggg\rangle \to |s3\rangle, |s4\rangle, |s5\rangle$ are blocked. The similar phenomena appear for other initial states such as $|gsgg\rangle$, $|ggsg\rangle$ and $|gggs\rangle$. Therefore, the four-qubit quantum phase gate can be also realized by using the above design.

The fidelity about three-qubit and four-qubit quantum phase gates can also be calculated similar to the two-qubit case. Figure 9 (a) and (b) show fidelities of the maximally entangled states as a function of distance for three- and four-qubit phase gates, respectively. The parameters of the metal sphere are taken identical with those in Fig.2. It is seen clearly that the value of fidelity changes with the distances between atoms and the metal sphere. The maximum of the fidelity appears at r=44nm for three-qubit case and r=53nm for four-qubit case. This means that the fidelities of three-qubit and four-qubit quantum phase gates can also be optimized by taking suitable distances between atoms and the metal sphere, which is similar to the case of two-quibt phase gate.

## 4. SUMMARY

Based on the Green's function approach, we have investigated the coupling of quantum dipole oscillators to the plasmon modes of the metal sphere. The Dicke subradiance and superradiance resulting from the interaction between surface plasmons of a nanosphere and an ensemble of quantum emitters have been analyzed. A scheme for the deterministic multi-qubit quantum gate has been proposed. As an example, two-qubit, three-qubit and four-qubit quantum phase gates have been designed in detail. The effect of losses in the metal on the efficiency of quantum gates has also been discussed. We would like to point out that our theory is suitable for designing any multi-qubit quantum phase gates although our discussions focus on two-, three- and four-qubit cases. Thus, the potential application of present phenomena to the quantum-information processing is anticipated.

**Acknowledgments**

This work was supported by the National Natural Science Foundation of China (Grant No. 11274042) and the National Key Basic Research Special Foundation of China under Grant 2013CB632704.

**Figure captions**

1. Fig.1 Plasmonic coupling of emitters near a metal nanoparticle.

2. Fig.2 (Color online) (a) Dispersion curves of n=1, 2, 3 surface polaritons for a metal sphere with different radii. (b) The corresponding extinction spectrum with a=20nm and r=25nm. The dielectric constant of the metal we use is described by Drude model, $\omega_p = 6.18\text{eV}$ and $\gamma = 0.05\text{eV}$.

3. Fig.3 (Color online) (a)The positions of two emitters (A and B) near a metal nanoparticle. (b) $\Gamma_{AB}/\Gamma$ as a function of the angle between two emitters around the sphere. The solid line, dashed line and dotted line correspond to the cases with $\omega = 0.55\omega_p, 0.625\omega_p$ and $0.652\omega_p$, respectively. The other parameters are identical with those in Fig.2.

4. Fig.4 Realization of a deterministic two-qubit quantum phase gate by applying external classical $2\pi$ pulses. (a) using $|gg\rangle$, subradiant and superradiant states. (b) for $|sg\rangle$, $|se\rangle$, $|gs\rangle$ and $|es\rangle$ states.

5. Fig. 5 (Color online) (a) $\Gamma_{AB}/\Gamma$ as a function of the angle between two emitters around the sphere. (b) Fidelity of a maximally entangled state created by the phase gate as a function of the distance between atoms and the sphere centre. (c) Coated sphere, a=20nm, b=24nm, and the parameters of the core are identical with those in Fig. 2. (d) The positions of the two atoms and the sphere used in (b). In (a) and (b), the lines in the same color represent the same system. Red: no coating; blue: coated with $\varepsilon' = 1.2-0.01i$; green: coated with $\varepsilon' = 1.2-0.05i$. The other parameters are identical with those in Fig.3.

6. Fig.6 Fidelity of a maximally entangled state created by the phase gate as a function of the distance between atoms and the centre of a metal aluminum sphere. The radii of aluminum core and coated sphere are taken as a=20nm and b=24nm, respectively. The red solid line: without coating at $\omega = 0.59\omega_p$, the dashed line and dotted line correspond to the aluminum sphere coated by $\varepsilon' = 1.2-0.1i$ and $\varepsilon' = 1.2-0.16i$ at $\omega = 0.57\omega_p$, respectively. The dielectric constant of aluminum is described by Drude model, $\omega_p = 2.27\times 10^4$ THz and $\gamma = 0.05\omega_p$.

7. Fig.6 Realization of a deterministic three-qubit quantum phase gate by applying external classical $2\pi$ pulses. (a) using $|ggg\rangle$, subradiant and superradiant states. (b) for $|gss\rangle, |sgs\rangle, |ssg\rangle$ states. (c) for $|sgg\rangle$ states.

8. Fig.7 Realization of a deterministic four-qubit quantum phase gate by applying external classical $2\pi$ pulses. (a) using $|gggg\rangle$, subradiant and superradiant states; (b) for $|gsss\rangle, |sgss\rangle, |ssgs\rangle$ and

|sssg⟩ states; (c) for |ssgg⟩ states; (d) for |sggg⟩ states.

9. Fig.8 (Color online) Fidelity of a maximally entangled state created by the phase gate as a function of the distance between the atoms and the sphere center. The inset shows the positions of atoms around the sphere. (a) for the case with three atoms; (b) for the case with four atoms. The other parameters are identical with those in Fig.2.

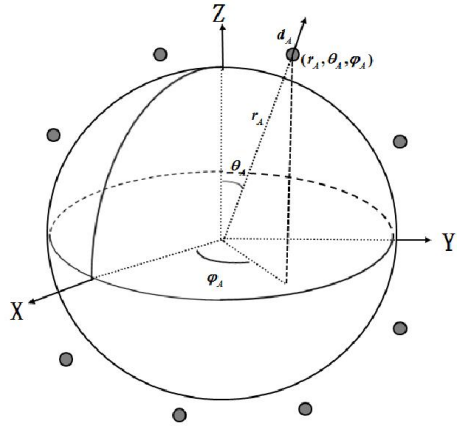

Fig.1

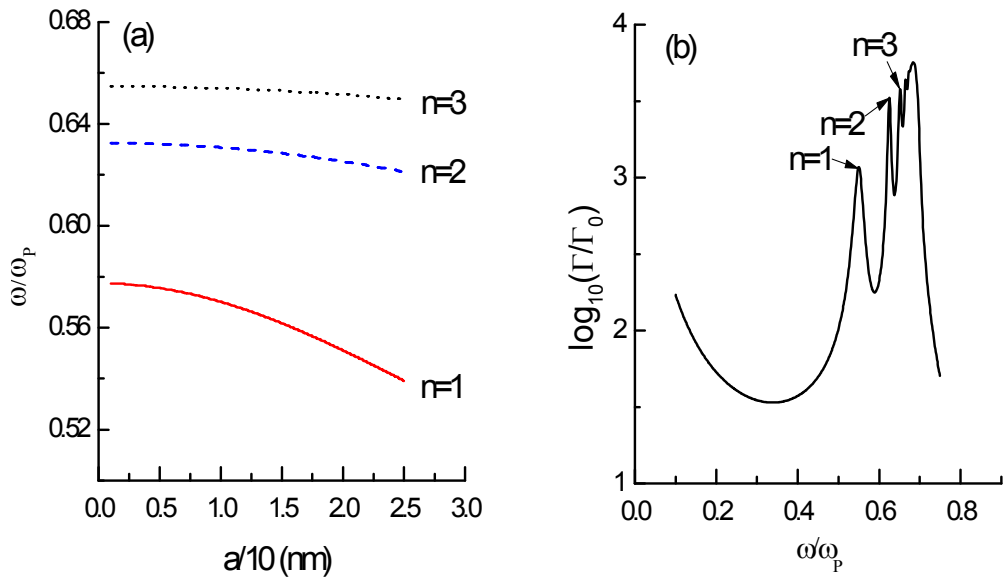

Fig.2

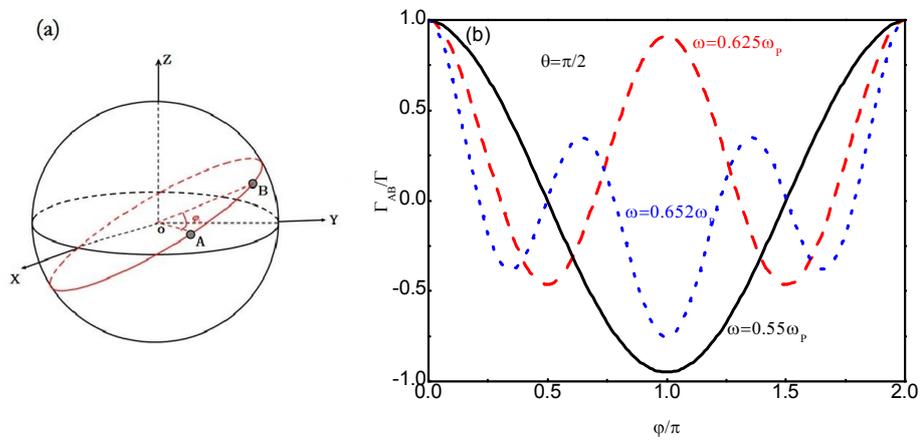

Fig.3

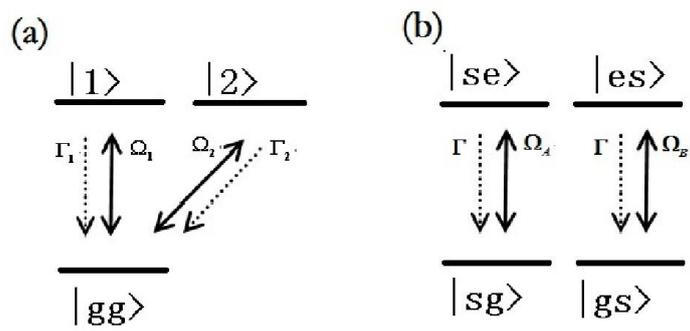

Fig.4

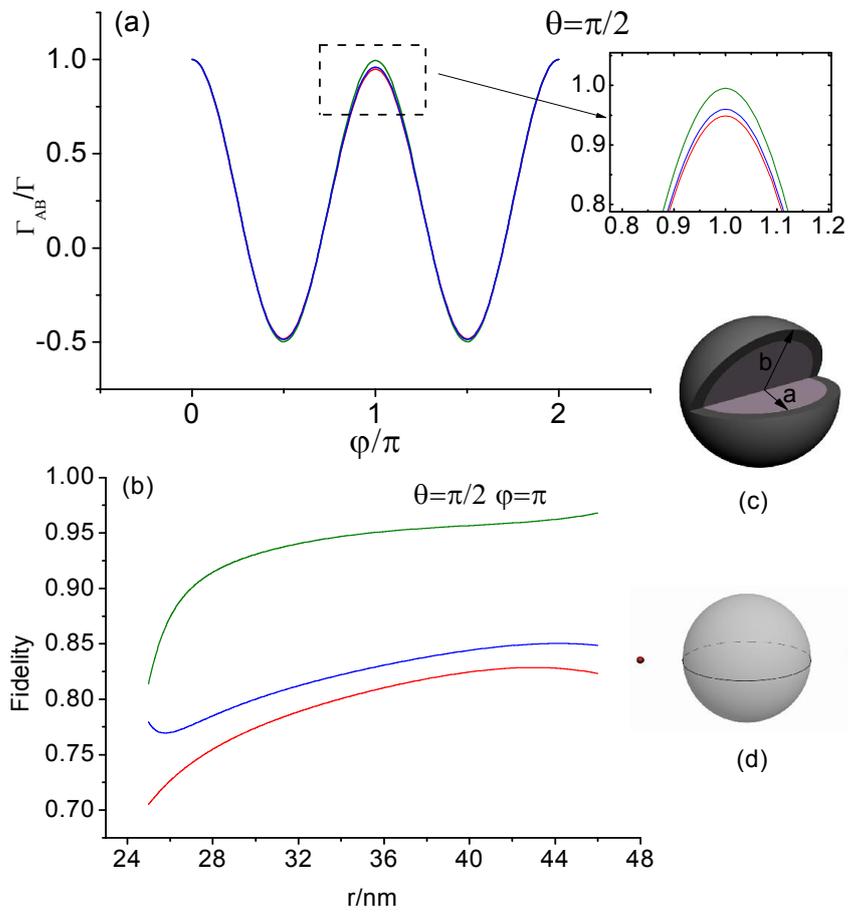

Fig.5

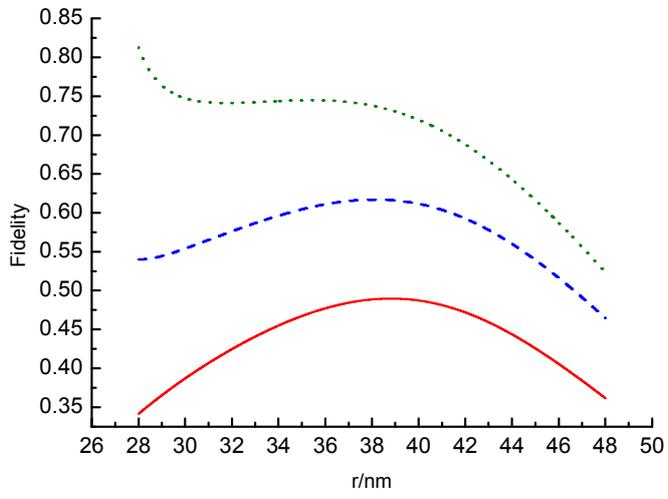

Fig.6

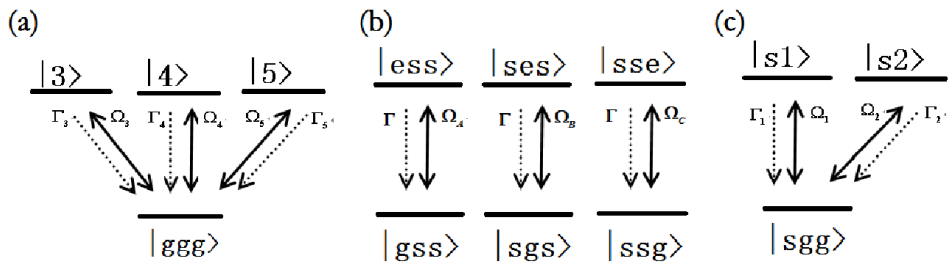

Fig.7

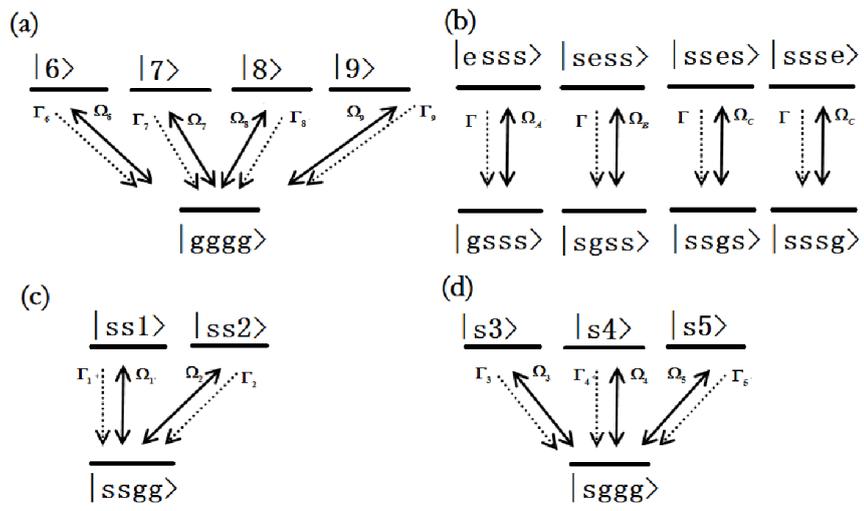

Fig.8

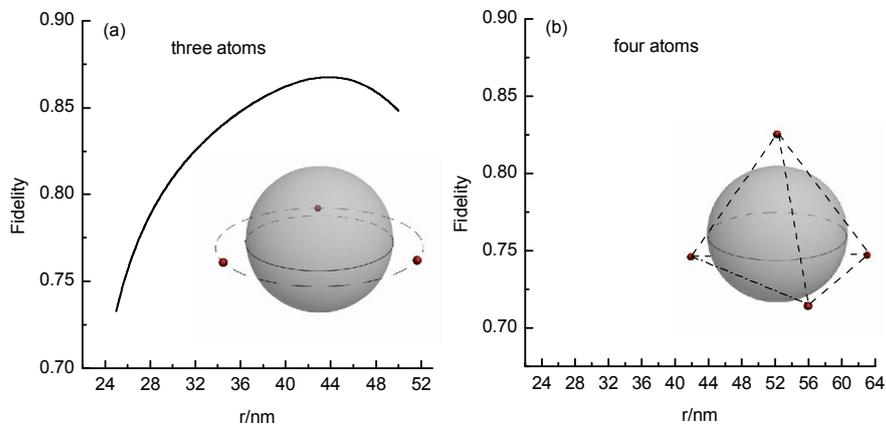

Fig.9